\documentclass[aps, a4paper,superscriptaddress, nofootinbib,12pt]{revtex4}
\usepackage{amsmath,amssymb, bm, natbib}
\usepackage{dsfont}
\usepackage{epsfig}
\usepackage{graphicx}
\usepackage{slashed}
\usepackage{color}

\usepackage{accents}
\usepackage{mathrsfs}
\usepackage[center,footnotesize,hang]{subfigure}
\usepackage{bigstrut}

\begin{document}

\title{Dirac or Inverse Seesaw Neutrino Masses from Gauged $B-L$ Symmetry}

\author{Ernest Ma}
\email{ma@phyun8.ucr.edu}
\affiliation{Department of Physics and Astronomy, University of California, \\
Riverside, California 92521, USA}
\author{Rahul Srivastava}
\email{rahuls@imsc.res.in}
\affiliation{The Institute of Mathematical Sciences, Chennai 600 113, India}


\begin{abstract}
   \vspace{1cm}
   
  \begin{center}
\textbf{ Abstract} 
\end{center}

The gauged $B - L$ symmetry is one of the simplest and well studied extension of standard model. In the conventional case, addition of three singlet right handed neutrinos each transforming as $-1$ under the $B - L$ symmetry renders it anomaly free.
 It is usually assumed that the $B - L$ symmetry is spontaneously broken by a singlet scalar having two units of $B - L$ charge, resulting in a natural implementation of Majorana seesaw mechanism for neutrinos. However, as we discuss in this proceeding, there is another simple anomaly free solution which leads to Dirac or inverse seesaw masses for neutrinos. These new possibilities are explored along with an application to neutrino mixing with $S_3$ flavour symmetry.

\end{abstract}

\maketitle


\section{Introduction}
\label{sec1}


  The nature of neutrinos i.e. whether they are Majorana or Dirac particles is one of the most important open questions in neutrino physics. Answering this question is essential to finding the underlying theory of neutrino masses and mixing. This issue can be potentially resolved by neutrinoless double beta decay experiments ($0\nu\beta\beta$). Currently several ongoing experiments are looking for signals of $0\nu\beta\beta$ but no such signal has been observed so far \cite{Auger:2012ar, Agostini:2013mzu, Gando:2012zm}. At present there is no compelling evidence from experiments or cosmological observations in favor of either Dirac or Majorana nature of neutrinos. With our current understanding, Dirac neutrinos are as plausible as Majorana ones. 

 However, theoretically Majorana neutrinos have received considerably more attention than Dirac neutrinos. There are several mechanisms (e.g. seesaw mechanisms) which satisfactorily explain smallness of neutrino masses if neutrinos are Majorana particles. On the other hand Dirac neutrinos are not so well studied. There are only few models capable of providing a natural explanation for smallness of Dirac neutrino masses. In this proceeding we present one simple model for Dirac neutrinos with naturally small masses based on gauged $B - L$ symmetry. This proceeding is based on our work \cite{Ma:2014qra} and interested reader is referred to it for further details.
 
 The plan of this proceeding is as follows. In Section \ref{sec2} we will briefly review the familiar scenario of gauged $B-L$ symmetry leading to the Majorana neutrinos and seesaw mechanism. The possibility of Dirac neutrinos in such a scenario will also be briefly discussed. In Section \ref{sec3} we will work out the new anomaly free solutions for gauged $B-L$ symmetry and show how it leads to Dirac neutrinos with naturally small masses. In Section \ref{sec4} we will expand on our results and using $S_3$ flavour symmetry will construct realistic neutrino mass matrices consistent with present oscillation data. In Section \ref{sec5} we will discuss the possibility of inverse seesaw Majorana neutrinos masses arising from the new anomaly free solutions. We will conclude in Section \ref{sec6}.


\section{Majorana neutrinos from $B - L$ Gauge Symmetry}
\label{sec2}


  Historically Baryon number ($B$) and Lepton numbers ($L_{i}$) were introduced to explain the stability of proton and absence of lepton flavour changing processes. In Standard Model (SM) the  baryon and lepton numbers turn out to be accidentally conserved classical symmetries. The $B$ and $L$ currents are anomalous and only the combination $B - L$ is anomaly free. These accidental symmetries need not be conserved by beyond standard model (BSM) physics. For example if one adds a Majorana mass term for neutrinos then the $B-L$ symmetry gets broken by 2 units. The addition of right handed neutrinos provides the possibility to promote this global $B-L$ symmetry to a anomaly free gauged $B - L$ symmetry. 

Now addition of any new $U(1)_x$ gauge symmetry implies that one needs to cancel anomalies \cite{Ma:2001kg, Ma:2002tc, Chen:2006hn}. The gauged $B-L$ symmetry can potentially induce both gauge as well as gauge-gravitational anomalies. The triangular gauge anomalies arising from gauged $U(1)_{B-L}$ are:
  
  \begin{itemize}
  
   \item $\rm{Tr} \left( U(1)_{B-L} \left[ SU(2)_L\right]^2 \right)$ 
   
   \item $\rm{Tr} \left( U(1)_{B-L} \left[ U(1)_Y\right]^2 \right)$
   
   \item  $\rm{Tr} \left( U(1)_{B-L} \right)^3$   
   
  \end{itemize}

    With the particle content of SM the first two anomalies are automatically canceled. Moreover, if the right handed neutrinos $\nu_{iR}$; $i= 1,2,3$ transform as $\nu_{iR} \sim -1$ under $U(1)_{B-L}$, then $\sum U(1)^3_{B-L} = 0$. The gauge - gravitational anomalies also vanish in this case as $-3 (-1) = 3$.  

Lets briefly discuss the possibility of Dirac neutrinos in this scenario before discussing the conventional case of Majorana neutrinos. The addition of right handed neutrinos with $-1$ charge under $U(1)_{B - L}$ allows one to have gauge invariant Yukawa coupling for neutrinos. The 
 $SU(3)_C \otimes SU(2)_L \otimes U(1)_Y \otimes U(1)_{B-L} $ invariant Yukawa coupling for neutrinos is then given by 
 
 \begin{eqnarray}
 - \mathcal{L}^\nu_Y = \sum_{i, j} \, y_{ij} \bar{L}_{iL} \hat{\Phi}^* \nu_{jR}  \, + \, \rm{h.c.}
  \end{eqnarray}

 where $\hat{\Phi}^* = i \tau_2 \Phi^*$ and $\Phi = (\phi^+ , \phi^0 )^{T}$ is the SM Higgs doublet. Since the right as well as left handed neutrinos transform non-trivially under the gauged $U(1)_{B-L}$ symmetry, this implies that the Majorana mass term for $\nu_{iR}$ is forbidden and neutrinos are Dirac particles. In this case the $U(1)_{B-L}$ symmetry remains unbroken \footnote{The $Z'$ gauge boson associated with $U(1)_{B-L}$ can get mass via Stuckelberg mechanism without breaking the $B-L$ symmetry \cite{Stueckelberg:1900zz, Feldman:2011ms, Heeck:2014zfa}.}.  However, in such a scenario the smallness of neutrino masses requires unnaturally small Yukawa couplings and the model does not provide any explanation for their smallness. For more details and variants of this scenario we refer to \cite{Heeck:2014zfa, Heeck:2013rpa, Heeck:2013vha}.

A relatively better understanding of smallness of neutrino masses can be obtained if in addition to the right handed neutrinos one also adds a singlet scalar $\chi$ transforming as  $\chi \sim 2$ under $U(1)_{B-L}$. In this case the $SU(3)_C \otimes SU(2)_L \otimes U(1)_Y \otimes U(1)_{B-L} $ invariant Yukawa coupling for neutrinos is given by 
 
 \begin{eqnarray}
 - \mathcal{L}^\nu_Y = \sum_{i, j} \, y_{ij} \bar{L}_{iL} \hat{\Phi}^* \nu_{jR}  \, + \,  \frac{1}{2} \sum_{i, j} \, f_{ij} \bar{\nu}^c_{iR} \chi \nu_{jR}   \, + \, \rm{h.c.}
  \end{eqnarray}
  
  The spontaneous symmetry breaking (SSB) of $\chi$ then leads to breaking of $B-L$ symmetry.  If $\langle \chi \rangle  = u $ the right handed neutrinos get a Majorana mass $M_{ij} = \sqrt{2} f_{ij} u $.
  If $u >> v$ i.e. the $B-L$ symmetry breaking scale is far greater than the electroweak scale, then $M_R >> m_D$ leading to a natural implementation of Type I seesaw mechanism.

  Before ending this section we like to remark that the gauged $B-L$ symmetry can be imbedded in other Beyond Standard Model scenarios e.g. it is an essential ingredient of Left-Right symmetric models. It can also be embedded in GUT groups e.g. $SO(10)$.


\section{Dirac Neutrinos from Gauged $B-L$ Symmetry}
  \label{sec3}


 In this section we look at the possibility of another simple choice of $B - L$ charges for right handed neutrinos which leads to anomaly free $U(1)_{B-L}$ gauge symmetry.  Unlike the previous case, let the 3 right handed neutrinos transform as $\nu_{iR} = (+5, -4, -4)$ under $B-L$ symmetry \cite{Ma:2014qra,Montero:2007cd, Machado:2010ui,Machado:2013oza}. Since $\nu_{iR} \sim (+5, -4, -4)$, therefore
 
 \begin{eqnarray}
  -(+5)^3 -(-4)^3  -(-4)^3    & = & +3 \\
   - (5) -(-4) - (-4)  & = & +3
 \label{rn}      
\end{eqnarray}

 Thus in this case also the model is free from gauge as well as gauge-gravitational anomalies.  Now the standard-model Higgs doublet $(\phi^+ , \phi^0 )^{T}$ does not connect $\nu_L$ with $\nu_R$. Therefore the neutrinos do not get mass from the standard electroweak symmetry breaking. To generate neutrino masses let us add three heavy Dirac singlet fermions $N_{L,R}$ transforming as $-1$ under $B-L$ symmetry.  They will not change the anomaly cancellation conditions and  the model will remain anomaly free. 
 Also let us add a singlet scalar $\chi_3$ transforming as $+3$ under $B - L$.

  Now for $\nu_{R2}$ and $\nu_{R3}$, $(\bar{\nu}_L, \bar{N}_L)$ is linked to $(\nu_R,N_R)$ through the $2 \times 2$ mass matrix  as follows

  \begin{equation}
M_{\nu,N} = \left( 
\begin{array}{cc}
0     & m_0  \\
m_3   & M  
\end{array}
\right)
 \label{dseesaw}
\end{equation}
 
 where $m_0$ comes from $\langle \phi^0 \rangle$. Moreover, $m_3$ comes from $\langle \chi_3\rangle$, due to the Yukawa coupling $\bar{N}_L \nu_R \chi_3$.  The invariant mass $M$ is naturally large, so the Dirac seesaw \cite{Roy:1983be} yields a small neutrino mass $m_3 m_0 /M$. 

  In the conventional $U (1)_{B-L}$ model, $\chi_2 \sim +2$ under $B-L$ is chosen to break the gauge symmetry, so that $\nu_R$ gets a Majorana mass and lepton number $L$ is broken to $(-1)^L$. Here, $\chi_3 \sim +3$ means that it is impossible to construct an operator of any dimension for a Majorana mass term and $L$ remains a conserved global symmetry, with $\nu_{L,R}$ and $N_{L,R}$ all having $L = 1$.
  
   Since $\nu_{R1} \sim +5$ does not connect with $\nu_L$ or $N_L$ directly, there is one massless neutrino in this case. The dimension-five operator $\bar{N}_L \nu_{R3}\chi^*_3 \chi^*_3/\Lambda$ is allowed by $U (1)_{B-L}$ and would give it a small Dirac mass.
    Alternatively, one can add a second scalar $\chi_6 \sim 6$ to the model to account for mass of $\nu_{R1}$.

 
\section{The $S_3$ Flavour Symmetry}
\label{sec4}


The discussion in previous section was aimed primarily at mass generation for Dirac neutrinos. The $U(1)_{B-L}$ symmetry alone does not provide any explanation for the currently observed PMNS mixing pattern.  In this section we will generalize our discussion and will construct phenomenologically viable lepton mass matrices. In order to understand the leptonic family structure consistent with present neutrino oscillation data, we will  make use of the non-Abelian discreet symmetry group $S_3$. 

 The $S_3$ group is the smallest non-Abelian discreet symmetry group and is the group of the permutation of three objects. It consists of six elements and is also isomorphic to the symmetry group of the equilateral triangle.  It admits three irreducible representations $1$, $1'$ and $2$ with the tensor product rules.

\begin{eqnarray}
  1 \otimes 1' & = & 1', \quad 1' \otimes 1' \, = \, 1, \quad 2 \otimes 1 \, = \, 2, \nonumber \\ 
2\otimes 1' & = & 2, \quad 2\otimes 2 \, = \, 1 \oplus 1' \oplus 2
 \label{prule}
\end{eqnarray}

  In this proceeding we will use the complex representation of the $S_3$ group \cite{Chen:2004rr,Ma:2013zca}. In the complex representation, if

 \begin{eqnarray}
 \left( \begin{array}{c}
\phi_1       \\
\phi_2     
\end{array} \right)\, , \quad
 \left( \begin{array}{c}
\psi_1       \\
\psi_2     
\end{array} \right)\,  \in \textbf{2}
\quad \Rightarrow \quad
\left( \begin{array}{c}
\phi^\dagger_2       \\
\phi^\dagger_1     
\end{array} \right)\, , \quad
 \left( \begin{array}{c}
\psi^\dagger_2        \\
\psi^\dagger_1     
\end{array} \right)\,  \in \textbf{2}
 \label{s3-2d-rep}
\end{eqnarray}
 then

 \begin{eqnarray}
  \phi_1 \psi_2 +  \phi_2 \psi_1 \, , \quad   \phi^\dagger_2 \psi_2 +  
\phi^\dagger_1 \psi_1  \quad \in \textbf{1}  \nonumber \\
  \phi_1 \psi_2 -  \phi_2 \psi_1 \, , \quad   \phi^\dagger_2 \psi_2 -  
\phi^\dagger_1 \psi_1  \quad \in \textbf{1}' \nonumber \\
  \left( \begin{array}{c}
\phi_2 \psi_2      \\
\phi_1 \psi_1    
\end{array} \right)\, , \quad
 \left( \begin{array}{c}
\phi^\dagger_1  \psi_2       \\
\phi^\dagger_2   \psi_1    
\end{array} \right)\,  \in \textbf{2}
   \label{s3tpro}
 \end{eqnarray}

  With this brief summary of $S_3$ group and its irreducible representations, we now move on to constructing an $S_3$ invariant lepton sector.   The $B-L$ charge and $S_3$ assignment of the fields for the lepton sector is as shown in Table \ref{tab1}.

\begin{table}[h]
\begin{center}
\begin{tabular}{c c c || c c c}
  \hline \hline
  Fields                      \hspace{0.4cm}                    & $B-L$                 \hspace{0.4cm}                                     &  $S_3$       \hspace{0.4cm}              & 
  Fields                      \hspace{0.4cm}                    & $B-L$                 \hspace{0.4cm}                                     &  $S_3$                                 \\
  \hline \hline
  $L^e$                       \hspace{0.4cm}                    & $-1$                   \hspace{0.4cm}                                    &  $1'$            \hspace{0.4cm}           &   
  $e_R$                       \hspace{0.4cm}                    & $-1$                    \hspace{0.4cm}                                   &  $1'$                                    \\
  $L^\mu$                     \hspace{0.4cm}                    & $-1$                   \hspace{0.4cm}                                    &  $1'$            \hspace{0.4cm}           &
  $\mu_R$                     \hspace{0.4cm}                    & $-1$                   \hspace{0.4cm}                                    &  $1'$                                    \\
  $L^\tau$                    \hspace{0.4cm}                    & $-1$                    \hspace{0.4cm}                                   &  $1$          \hspace{0.4cm}           &   
  $\tau_R$                    \hspace{0.4cm}                    & $-1$                    \hspace{0.4cm}                                   &  $1$                                    \\
  $N^1_L $                    \hspace{0.4cm}                    & $-1$                    \hspace{0.4cm}                                   &  $1'$               \hspace{0.4cm}            &
  $N^1_R $                      \hspace{0.4cm}                  & $-1$                     \hspace{0.4cm}                                  &  $1'$                                   \\ 
  $N^2_L $                     \hspace{0.4cm}                   & $-1$                     \hspace{0.4cm}                                  &  $1'$               \hspace{0.4cm}        & 
  $N^2_R $                      \hspace{0.4cm}                  & $-1$                     \hspace{0.4cm}                                  &  $1'$                                   \\
  $N^3_L $                      \hspace{0.4cm}                  & $-1$                    \hspace{0.4cm}                                   &  $1$                   \hspace{0.4cm}           & 
  $N^3_R $                        \hspace{0.4cm}                & $-1$                     \hspace{0.4cm}                                  &  $1$                                     \\
   $\Phi$                       \hspace{0.4cm}                  & $0$                      \hspace{0.4cm}                                  & $1$           \hspace{0.4cm}               &
  $\nu^e_R $                   \hspace{0.4cm}                   & $5$                      \hspace{0.4cm}                                  &  $1'$                                    \\  
       $ \left( \begin{array}{c}
  \nu^\mu_R   \\ \nu^\tau_R   
  \end{array} \right)$               \hspace{0.4cm}               & $ -4$                     \hspace{0.4cm}                                 &  $2$            \hspace{0.4cm}           &    
  \vspace{0.5cm}
  $ \left( \begin{array}{c}
  \chi_2   \\ \chi_3                                                           
  \end{array} \right)$               \hspace{0.4cm}               & $ 3$                         \hspace{0.4cm}                             &  $2$                                    \\
  \hline
  \end{tabular}
\end{center}
\caption{The $B-L$ and $S_3$ charge assignment for the fields.}
  \label{tab1}
\end{table}    

 where  we denote the left handed lepton doublets by $L^\alpha = \left( \nu^\alpha_L , l^\alpha_L  \right)^{\rm{T}}$ where $\alpha = e, \mu, \tau$; the right handed charged leptons are denoted as $e_R, \mu_R, \tau_R$ and the right handed neutrinos 
as $\nu^{e}_R, \nu^{\mu}_R, \nu^{\tau}_R$. Also, let us denote the  heavy singlet fermions as $N^i_{L,R}$; $i=1,2,3$. The ``Standard Model like'' scalar doublet is denoted by $\Phi = \left( \phi^+ , \phi^0  \right)^{\rm{T}}$ and the singlet scalars are denoted by $\chi_{2,3} $.

    The $S_3$ and $B-L$ invariant Yukawa interaction $\mathcal{L}_Y$ can then be written as 
    
  \begin{eqnarray}
  \mathcal{L}_Y & = &    \mathcal{L}_{L^\alpha l_R} }  \, + \,   
\mathcal{L}_{L^\alpha N_R}   \, + \,  \mathcal{L}_{N_L N_R}      \, + \,  
\mathcal{L}_{N_L \nu_R
 \label{yuk}
\end{eqnarray}

 where 
\begin{eqnarray}
\mathcal{L}_{L^\alpha l_R}  & = &  y'_e \, \bar{L}^e \, \Phi \, e_R    \, 
+ \, y'_{12} \, \bar{L}^e \, \Phi \, \mu_R    \, + \, y'_{21} \, \bar{L}^\mu \, 
\Phi \, e_R    \, + \, y'_\mu \, \bar{L}^\mu \, \Phi \, \mu_R  \, + \,  y_\tau \,  \bar{L}^\tau \, \Phi \, \tau_R        \nonumber \\
\mathcal{L}_{L^\alpha N_R}  & = &  g'_{11} \, \bar{L}^e \, \hat{\Phi}^* \, N^1_R    
\, + \, g'_{12} \, \bar{L}^e \, \hat{\Phi}^* \, N^2_R    \, + \, g'_{21} \, 
\bar{L}^\mu \, \hat{\Phi}^* \, N^1_R    \, + \,  g'_{22} \, \bar{L}^\mu \, 
\hat{\Phi}^* \, N^2_R  \nonumber \\
& + &  g_{33}  \, \bar{L}^\tau \, \hat{\Phi}^* \, N^3_R     \nonumber \\
\mathcal{L}_{N_L N_R}  & = & M'_{11} \, \bar{N}^1_L \, N^1_R  \, + \, M'_{12} \, 
\bar{N}^1_L \, N^2_R  \, + \, M'_{21} \, \bar{N}^2_L \, N^1_R  \, + \,  M'_{22} \, 
\bar{N}^2_L \, N^2_R \, + \,  M_{33} \, \bar{N}^3_L  \, N^3_R   \nonumber \\
\mathcal{L}_{N_L \nu_R}  & = &   \frac{f'_{11}}{\Lambda} \, \left (\bar{N}^1_L \, 
\nu^e_R \right) \otimes \left[ \left( \begin{array}{c}  \chi^*_3   \\ 
\chi^*_2   \end{array} \right) \, \otimes \, \left( \begin{array}{c}  
\chi^*_3   \\ \chi^*_2  \end{array} \right) \right]_{1}     
\, + \, \frac{f'_{21}}{\Lambda} \, \left (\bar{N}^2_L \, \nu^e_R \right) 
\otimes \left[ \left( \begin{array}{c}  \chi^*_3   \\ \chi^*_2   \end{array} 
\right) \, \otimes \, \left( \begin{array}{c}  \chi^*_3   \\ \chi^*_2  
\end{array} \right) \right]_{1} 
\nonumber \\
& + &  f'_{12} \, \bar{N}^1_L \, \otimes \, \left[  \left( \begin{array}{c}  
\nu^\mu_R   \\ \nu^\tau_R  \end{array} \right) \, \otimes \, \left( 
\begin{array}{c}  \chi_2   \\ \chi_3   \end{array} \right) \right]_{1'}     
\, + \, f'_{22} \, \bar{N}^2_L \, \otimes \, \left[  \left( \begin{array}{c}  
\nu^\mu_R   \\ \nu^\tau_R  \end{array} \right) \, \otimes \, \left( 
\begin{array}{c}  \chi_2   \\ \chi_3   \end{array} \right) \right]_{1'} 
\nonumber \\
& + &  f_{33} \, \bar{N}^3_L \, \otimes \, \left[  \left( \begin{array}{c}  
\nu^\mu_R   \\ \nu^\tau_R  \end{array} \right) \, \otimes \, \left( 
\begin{array}{c}  \chi_2   \\ \chi_3   \end{array} \right) \right]_{1}  
\label{cly}
\end{eqnarray}

  In writing (\ref{cly}), we have used the notation $\hat{\Phi}^* = i \tau_2 \Phi^* = (\phi^0, \phi^-)^{\rm{T}}$.  Here, $y_\alpha$ are the Yukawa couplings of the charged leptons whereas  $f_{ij}$, $g_{ij}$ and $M_{ij}$ denote the dimensionless coupling constants between the leptons and the heavy fermions.  
   
        At this point we like to remark that in $\mathcal{L}_Y$ there is still a freedom to redefine a few fields (i.e. the pairs $N^1_{L} - N^2_L$, $N^1_R - N^2_R$ and $e_R - \mu_R$) in a way that certain couplings can be made equal to zero.
   For sake of later convenience we choose to use this freedom of field redefinition to make $f'_{11} = M'_{12} = y'_{21} =0$. Moreover, we relabel the remaining non-zero couplings of these redefined fields as $f'_{ij} \to f_{ij}, g'_{ij} \to g_{ij}, M'_{ij} \to M_{ij}$. 
   
   After symmetry breaking the scalar fields get VEVs $\langle \phi^0 \rangle = v$, $\langle \chi_i \rangle = u_i$; $i = 2,3$.  Then the mass matrix relevant to charged leptons is given by 

 \begin{eqnarray}
   \mathcal{M}_l =   v \, \left( \begin{array}{c c c} 
                      y_e     &  y_{12}    &  0     \\
                      0         &  y_\mu       &  0      \\ 
                      0         &  0          &   y_\tau
                      \end{array} \right) 
  \label{lm}
 \end{eqnarray}

  This mass matrix can be readily diagonalized by bi-unitary transformation. In the limit of $y_e << y_\mu$ we get 

  \begin{eqnarray}
    \theta^l_{12} & \approx & \tan^{-1}\left( \frac{-y_{12}}{y_\mu} \right); 
\qquad \qquad \quad  m_e \approx v \, y_e \rm{cos} \, \theta^l_{12}   
\nonumber \\
    m_\mu & \approx & v \left( y_\mu \rm{cos} \, \theta^l_{12} - y_{12} 
\rm{sin}\, \theta^l_{12}   \right) ; \qquad m_\tau \approx v y_\tau
   \label{lmu}
  \end{eqnarray}
 
   If  $y_{12} = y_\mu$ then maximal mixing is achieved i.e. $ \theta^l_{12} = -\frac{\pi}{4}$, with $m_\mu = \sqrt{2} v y_\mu$. 
   
   Also, the $6 \times 6$ mass matrix spanning $(\bar{\nu}^e_L, \bar{\nu}^\mu_L, \bar{\nu}^\tau_L, \bar{N}^1_L, \bar{N}^2_L, \bar{N}^3_L)$ and $(\nu^e_R, \nu^\mu_R, \nu^\tau_R, N^1_R, N^2_R, N^3_R)^{\rm{T}}$ of neutrinos and the heavy fermions is given by

\begin{eqnarray}
   \mathcal{M}_{\nu,N} =    \, \left( \begin{array}{c c  c c c c} 
                      0                           &   0                  & 0                  &   g_{11} v^*        &    g_{12} v^*             &   0                       \\ 
                      0                           &   0                  &  0                 &   g_{21} v^*        &    g_{22}  v^*            &   0                        \\ 
                      0                           &   0                  &  0                 &   0                 &    0                      &   g_{33}  v^*              \\ 
                      0                           &   f_{12} u_3         & -f_{12} u_2        &  M_{11}             &    0                      &   0                         \\ 
         \frac{f_{21}}{\Lambda} u^*_2 u^*_3       &   f_{22} u_3         & -f_{22} u_2        &  M_{21}             &    M_{22}                 &   0                         \\ 
                      0                           &   f_{33} u_3         &   f_{33} u_2       &  0                  &    0                      &  M_3 
                      \end{array} \right) 
  \label{lnuN}
 \end{eqnarray}
 
        As remarked earlier, the mass terms $M_{ij}$ between the heavy fermions can be naturally large, so we can block diagonalize the mass matrix assuming that $f_{ij}, g_{ij} << M_{ij}$. The block diagonalized mass matrix of light neutrinos is given by
 
 \begin{scriptsize}
 \begin{eqnarray}
 && \mathcal{M}_\nu   =   m_{N_L \nu_R} \left(M_{N_L N_R}\right)^{-1} m_{L^\alpha N_R}  \nonumber \\
 & & =   v^*  \left( \begin{array}{c c c} 
 \frac{(g_{21} M_{11} - g_{11} M_{21} )f_{12} u_2}{M_{11} M_{22}}                                                                &        \frac{(g_{22} M_{11} - g_{12} M_{21}) f_{12} u_2}{M_{11} M_{22}}               
            &             \frac{-f_{12} g_{33} u_3}{M_{33}}                  \\
 \frac{(g_{21} M_{11} - g_{11} M_{21} )f_{22} u_2 \, +\, f_{21} g_{11} M_{22} u_6}{M_{11} M_{22}}        &      \frac{(g_{22} M_{11} - g_{12} M_{21}) f_{22} u_2 \, +\, f_{21} g_{12} M_{22} u_6}{M_{11} M_{22}}    
            &            \frac{-f_{22} g_{33} u_3}{M_{33}}     \\ 
\frac{(g_{21} M_{11} - g_{11} M_{21} )f_{33} u_2}{M_{11} M_{22}}                                                                 &       \frac{(g_{22} M_{11} - g_{12} M_{21}) f_{33} u_2}{M_{11} M_{22}}        
            &             \frac{f_{33} g_{33} u_3}{M_{33}} 
                      \end{array} \right)  \nonumber \\
  \label{nu}
 \end{eqnarray}
 \end{scriptsize}
 
   where we have written $u_6 = \frac{u^*_2 u^*_3}{\Lambda}$.  Also, the $3 \times 3$ mass matrices $m_{L^\alpha N_R}$, $M_{N_L N_R}$ and $m_{N_L \nu_R}$ are obtained from the terms $\mathcal{L}_{L^\alpha N_R}$, $\mathcal{L}_{N_L N_R}$ and $\mathcal{L}_{N_L \nu_R}$
respectively. This light neutrino mass matrix can be further diagonalized by the bi-unitary transformation. 

  The neutrino masses and the mixing angles so obtained will be dependent on the specific values of the coupling constants $f_{ij}, g_{ij}, M_{ij}$  as well as the VEVs $v, u_i$; $i = 2,3$.  In the simplifying case of $g_{ij} = g$ and $M_{ij} = M$ we get 
  
\begin{eqnarray}
 \mathcal{M}_\nu =   \frac{g v^*}{M}  \left( \begin{array}{c c c} 
               0             &         0                       &             -f_{12}  u_3                  \\
          f_{21}  u_6        &     f_{21}  u_6                 &            -f_{22}  u_3     \\ 
               0             &         0                       &             f_{33} u_3
                      \end{array} \right) 
  \label{nus}
\end{eqnarray}

   Diagonalizing the mass matrix we have 

  \begin{eqnarray}
    \theta^\nu_{12} & \approx & 0 ;  \quad   \theta^\nu_{13}  \approx   \tan^{-1}\left( \frac{f_{12}}{f_{33}}  \right);     \quad    \theta^\nu_{23}  \approx   \tan^{-1}\left( \frac{f_{22}}{\sqrt{f^2_{12} + f^2_{33}}}  \right)    
    \nonumber \\
   m^\nu_1 & \approx & 0 ; \quad  \quad  m^\nu_2  \approx  \frac{\sqrt{2(f^2_{12} + f^2_{33})} f_{21} g |v|}{M \sqrt{f^2_{12} + f^2_{22} + f^2_{33}}}\, |u_6|; \nonumber \\
      m^\nu_3 & \approx  &  \frac{\sqrt{f^2_{12} + f^2_{22} + f^2_{33}}\, g |v|}{M }\, |u_3|
   \label{numu}
  \end{eqnarray}

    Since, $u_6 << u_3$, we have a normal hierarchy pattern with two nearly massless neutrinos and one relatively heavy neutrino. Moreover, the massless neutrino will also gain small mass,  if any of the $M_{ij}$'s or $g_{ij}$'s are not equal to $M$ or $g$  respectively. Also, if they deviate significantly from these values then one can possibly recover degenerate or inverted hierarchy patterns also. Now, if $U_{l}$ and $U_\nu$ are the mixing matrices of the charged leptons and neutrinos respectively, 
    then the PMNS mixing matrix is given by 

\begin{eqnarray}
  U_{\rm{PMNS}}  = U^\dagger_{l} U_\nu
 \label{pmns}
\end{eqnarray}

    Taking $y_{12} = y_\mu$, $f_{12} = -\frac{f_{33}}{2}$ and $f_{22} = \sqrt{f^2_{12} + f^2_{33}}$ in (\ref{lmu}), (\ref{numu}) we get $\theta^\nu_{23} = -\theta^l_{12} =  \frac{\pi}{4}$ and $\theta^\nu_{13} = \tan^{-1}(-\frac{1}{2})$ which gives PMNS mixing angles consistent with present $3-\sigma$ limits of global fits obtained from experiments \cite{Capozzi:2013csa}.

 In our minimal model with only one doublet scalar, the quark sector can be accommodated in a simple way if both the left handed quark doublets $Q^i_L = (u^i_L, d^i_L)^{\rm{T}}$, $i= 1,2,3$ and the right handed quark singlets $u^i_R$, $d^i_R$; $i=1,2,3$
transform as $1$ of $S_3$.  A better understanding of the quark sector can be obtained if, to our minimal model, we add more doublet scalars transforming non-trivially under $S_3$.  One such example for quark sector, albeit in context of a 
different model for lepton sector, has already been worked out in \cite{Chen:2004rr, Ma:2013zca}. We are currently working on a similar extension of our minimal model.


\section{Inverse Seesaw }
 \label{sec5}


 Apart from the Dirac neutrinos, other possibilities can also arise depending on the particle content \cite{Ma:2014qra, Machado:2013oza}.  For example instead of adding the singlet scalar $\chi_3 \sim 3$ under $U(1)_{B-L}$, one can add two complex scalar fields $\chi_2 \sim 2$ and $\chi_6 \sim 6$ under $U (1)_{B-L}$. In this case, $\nu_L$ is not connected to $\nu_{R1,R2} \sim -4$. It is connected however to $N_{L,R}$ through the mass matrix spanning $(\bar{\nu}_L, \bar{N}^c_R, \bar{N}_L)$ as follows:

  \begin{equation}
M_{\nu,N} = \left( 
\begin{array}{ccc}
0     & m_0     & 0 \\
m_0   & m'_2   & M \\
0     & M       & m_2
\end{array}
\right)
 \label{invseesaw}
\end{equation}

where $m_2$ and $m'_2$ come from the Yukawa couplings with $\chi_2$. This leads to an inverse seesaw \cite{Wyler:1982dd, Mohapatra:1986bd, Ma:1987zm}, i.e. $m_\nu \simeq m^2_0 m_2/M^2$. In the case of $\nu_{R3} \sim +5$, the corresponding mass matrix spanning $( \bar{\nu}_L, \bar{N}^c_R, \bar{N}_L,  \bar{\nu}^c_{R3})$ is given by

 \begin{equation}
M_{\nu,N} = \left( 
\begin{array}{cccc}
0     & m_0     & 0    & 0    \\
m_0   & m'_2    & M    & 0    \\
0     & M       & m_2  & m_6   \\
0     & 0       & m_6  & 0
\end{array}
\right)
 \label{lopseesaw}
\end{equation}
where $m_6$ comes from the Yukawa coupling with $\chi_6$.  Thus $\nu_{R3}$ also gets an inverse seesaw mass $\simeq m^2_6 m'_2 / M^2$ which is the $4 \times 4$ analog of the $3 \times 3$ lopsided seesaw discussed in \cite{Ma:2009du}. In this scheme, 
$\nu_L$ and $\nu_{R3}$ get small masses via inverse seesaw mechanism. Also, $N_{1,2,3}$ become heavy pseudo-Dirac fermions. However, $\nu_{R1,R2}$ remain massless.  They can be given mass by adding extra scalars e.g. by adding a third scalar 
$\chi_8 \sim 8$ under $B - L$.


\section{Conclusion and Future Work}
 \label{sec6}


 The idea that $B - L$ should be a gauge symmetry has been around for some time.  However most of the work on gauged $B - L$ symmetry has been done for the case of the three right handed neutrinos transforming as
- 1 under $U(1)_{B-L}$.  In this work we looked at another possible anomaly free solution for gauged $B - L$ interaction with the three right handed neutrinos transforming as (+ 5, - 4, - 4) under $U(1)_{B-L}$. 
   We showed how these assignments can be used to obtain seesaw Dirac neutrino masses, as well as inverse seesaw Majorana neutrino masses.  We then showed that imposition of $S_3 $ flavour symmetry to the first case can lead to realistic neutrino and charged-lepton mass matrices with a mixing pattern consistent with experiments. In our model the $B - L$ symmetry breaking scale can be as low as in TeV range.  This raises the possibility of testing it in future runs of LHC or in other future colliders. 
 We are planning to look for the phenomenological consequences of our model. The phenomenology of the $Z'$ boson would be of particular interest. The cosmological implications of our model will also be interesting.


\begin{acknowledgments}

  RS will like to thank the organizers for inviting and giving opportunity to present this work at  International Conference on Massive Neutrinos - 2015, Singapore. EM is supported in part by the U. S. Department of Energy under 
Grant No. DE-SC0008541.

\end{acknowledgments}



\end{document}